\documentclass[manuscript,nonacm]{acmart}

\usepackage{amsmath}
\usepackage{float}
\usepackage{graphicx}
\usepackage{subfig}

\usepackage{enumitem}
\usepackage{array}
\newcolumntype{H}{>{\setbox0=\hbox\bgroup}c<{\egroup}@{}} 

\usepackage{url}
\AtBeginDocument{%
  \providecommand\BibTeX{{%
    \normalfont B\kern-0.5em{\scshape i\kern-0.25em b}\kern-0.8em\TeX}}}

\setcopyright{none}

\renewcommand\footnotetextcopyrightpermission[1]{} 
\makeatletter
\let\@authorsaddresses\@empty
\makeatother

\begin{document}

\fancyhead{}

\title{On the challenges of studying bias in Recommender Systems: A UserKNN case study}
\titlenote{Copyright 2024 for this paper by its authors. Use permitted under Creative Commons License Attribution 4.0 International (CC BY 4.0).\newline Accepted at FAccTRec 2024.}

\author{Savvina Daniil}

\email{s.daniil@cwi.nl}
\affiliation{%
  \institution{Centrum Wiskunde \& Informatica}
  \country{Amsterdam, The Netherlands}
}

\author{Manel Slokom}
\email{m.slokom@cwi.nl}
\affiliation{%
  \institution{Centrum Wiskunde \& Informatica}
  \country{Amsterdam, The Netherlands}
}

\author{Mirjam Cuper}
\email{mirjam.cuper@kb.nl}
\affiliation{%
  \institution{National Library of the Netherlands}
  \country{The Hague, The Netherlands}
}
\author{Cynthia C.~S. Liem}
\email{c.c.s.liem@tudelft.nl}
\affiliation{%
  \institution{Delft University of Technology}
  \country{Delft, The Netherlands}
}
\author{Jacco van Ossenbruggen}
\email{jacco.van.ossenbruggen@cwi.nl}
\affiliation{%
  \institution{Vrije Universiteit Amsterdam}
  \country{Amsterdam, The Netherlands}
}
\author{Laura Hollink}
\email{l.hollink@cwi.nl}
\affiliation{%
  \institution{Centrum Wiskunde \& Informatica}
  \country{Amsterdam, The Netherlands}
}

\renewcommand{\shortauthors}{Daniil}

\begin{abstract}
Statements on the propagation of bias by recommender systems are often hard to verify or falsify.
Research on bias tends to draw from a small pool of publicly available datasets and is therefore bound by their specific properties.
Additionally, implementation choices are often not explicitly described or motivated in research, while they may have an effect on bias propagation.
In this paper, we explore the challenges of measuring and reporting popularity bias.
We showcase the impact of data properties and algorithm configurations on popularity bias by combining synthetic data with well known recommender systems frameworks that implement UserKNN.
First, we identify data characteristics that might impact popularity bias, based on the functionality of UserKNN. 
Accordingly, we generate various datasets that combine these characteristics.
Second, we locate UserKNN configurations that vary across implementations in literature.
We evaluate popularity bias for five synthetic datasets and five UserKNN configurations, and offer insights on their joint effect.
We find that, depending on the data characteristics, various UserKNN configurations can lead to different conclusions regarding the propagation of popularity bias.
These results motivate the need for explicitly addressing algorithmic configuration and data properties when reporting and interpreting bias in recommender systems.
\end{abstract}



\maketitle

\section{Introduction}
Recommender systems are commonly used as a tool to encode taste based on the information available, be it user history or metadata.
The wide use of recommender systems necessitates critical reflection on the issues that may arise when we allow automation to dictate our exposure to information.
Specifically, bias in recommender systems is a topic of interest within the scholarly community. 
Bias is a complex term that can refer to various types of biases associated with interactions between users and items in a given system \cite{10.1145/3564284}.

Many studies have focused on specifically measuring the phenomenon of \textit{popularity bias} in collaborative filtering systems \cite{ahanger2022popularity, elahi2021investigating, yalcin2021blockbuster, abdollahpouri2020popularity, zhao2022popularity}.
Despite this large research effort to track and mitigate popularity bias, there is no univocal message regarding why and when it occurs. 
Studies that measure popularity bias propagated by commonly used algorithms on benchmark datasets report varying, sometimes contradicting results \cite{daniil2024reproducing}.
This observation raises questions; is popularity bias sensitive to properties of the system that do not receive sufficient attention?
Why is a seemingly simple phenomenon so hard to study?
We hypothesize that two factors complicate bias measuring and reporting, namely data characteristics and algorithm configurations.

\paragraph{Data characteristics}
Benchmark datasets are very useful for academic research, as they allow researchers to evaluate their hypotheses and benchmark their proposed debiasing methods. 
However, their consistent use raises concerns that relate to the dependence on the domain and source they were constructed from, and the potentiality for blind spots that stem from outdated rating behaviour. 
Most importantly, by reporting on different types of bias such as popularity bias on only a small set of publicly available datasets \cite{wang2023survey}, researchers are restricted by their specific characteristics. 
This specificity limits the scope of research \cite{cremonesi2021progress}, and obfuscates the process of examining causality. 
In other words, it is not trivial to conclude whether the propagation of bias or lack thereof is a result of the respective algorithm's functionality, or of certain intricate details of the user-item interactions within these datasets. 
Informed data synthesis can potentially assist with overcoming this barrier and gaining a holistic view of bias propagation by recommender systems. 

\paragraph{Algorithm configuration}
Insufficient reporting of algorithm configurations leads to a reproducibility problem within research on recommender systems.
Studies have shown that papers published in top-tier conferences often do not disclose sufficient information for replication and verification \cite{ferrari2021troubling}.
This issue is also relevant in the bias discussion. 
Even relatively simple algorithmic approaches, such as neighbour-based ones, are constructed using hyperparameters and implementation choices that might affect whether bias propagation is observed. 
The RecSys community proposes a set of evaluation frameworks to promote reproducibility\footnote{\url{https://github.com/ACMRecSys/recsys-evaluation-frameworks}}, but we found that there are important differences between them that often go unmentioned. 
Testing the effect of algorithm configuration can be a means of reporting on bias in a comprehensive manner.

\vspace{0.25cm}

In this paper, we experiment with data characteristics and algorithm configurations and observe the effect on popularity bias, with a focus on UserKNN.
First, we look into data characteristics that might have an impact on popularity bias given a rating prediction and top-10 recommendation task, a common setup among recent studies on popularity bias \cite{abdollahpouri2019unfairness,abdollahpouri2020connection,kowald2020unfairness}.
Specifically, we delve into the relation between popularity and rating, as well as the preferences of users with large profiles.
We form a set of data scenarios by tweaking and combining these characteristics.
For each scenario, we generate a corresponding synthetic dataset of ratings, based on the interactions from a subset of Book-Crossing \cite{ziegler2005improving} constructed by \cite{naghiaei2022unfairness}.
Second, for UserKNN we identify configuration choices that may impact whether or not popularity bias is observed, and often differ across implementations in frameworks recommended by ACM RecSys, with LensKit \cite{ekstrand2020LensKit} and Cornac \cite{salah2020cornac} being tested in this paper. 
Despite the simplicity of UserKNN, there are certain configuration choices that can potentially greatly influence the result. 
We perform the recommendation process for each synthetic dataset with varied UserKNN configuration choices.
We apply commonly used popularity bias metrics to evaluate the recommended lists, as well as RMSE and NDCG@10 to estimate the performance when it comes to rating prediction and ranking.

Our results show that popularity bias is not always present in the recommended lists.
Whether popularity bias is observed, and to what extent, depends on the characteristics of the dataset and configuration choices of UserKNN, which also relate to the framework implementation.
The relationship between rating and popularity, as well as the preferences of users with large profiles are crucial when it comes to popularity bias propagation.
Additionally, UserKNN configurations such as minimum similarity and minimum neighbours largely impact the intensity of popularity bias.

The contributions of this paper are as follows:
\begin{itemize}[topsep=3pt]
\item a systematic investigation into the effect of data characteristics on popularity bias, by comparing results on five synthetic datasets for which we control the properties.  
\item a systematic investigation into the effect of implementation differences, by comparing results of UserKNN configurations as well as non-configurable implementation differences in well known frameworks.
\item we highlight standout results among the many, to give insights into why certain combinations of dataset characteristics and UserKNN configurations lead to popularity bias. 
\end{itemize}

With this work, we wish to contribute to the field by highlighting and disentangling the challenges in studying popularity bias in recommender systems.

\section{Related Work}
In this section, we provide a brief overview of existing work on bias in recommender systems and datasets and reproducibility.
\subsection{Bias in Recommender Systems}

Recommender systems are not immune to bias, even when only user consumption history is fed to the model and not other information about the users or items. 
\cite{test} discuss that a model might learn sensitive information like the gender of the user in the latent space, and produce recommendations that are gender-dependent, even more so than the interactions observed in the training set itself. 
In a survey on the topic of bias and debias in recommender systems research, \cite{10.1145/3564284} identify three factors that contribute to bias: user behavior's dependence on the exposure mechanism of the system, imbalanced presence of items (and users) in the data, and the effect of feedback loops. 
One type of bias that arises from the interaction between an algorithm and imbalanced data is popularity bias.

Popularity bias is the phenomenon where popular items (i.e., items that are frequently interacted with in the dataset) are recommended even more frequently than their popularity would warrant \cite{abdollahpouri2020multisided}. 
It is commonly believed to be caused by the long-tail distribution that often characterizes user-item interactions: most items have been rated by only a few users, and a few items have been rated by many users \cite{brynjolfsson2006niches}.
Various studies have reported that frequently used recommender systems algorithms are prone to propagating popularity bias existing in the dataset they were trained on \cite{abdollahpouri2019unfairness,kowald2020unfairness,naghiaei2022unfairness}. 
Different metrics have been proposed to quantify popularity bias \cite{abdollahpouri2019unfairness,abdollahpouri2017controlling}. 
Despite the extensive literature, our understanding of why certain algorithms and datasets are more or less prone to popularity bias is limited.
Additionally, studies that measure popularity bias on the same algorithms and datasets sometimes report conflicting results \cite{daniil2024reproducing}.
In this paper, we describe specific scenarios of data-algorithm interaction and report the results of different metrics associated with popularity bias.

\subsection{Datasets and Reproducibility}
The way that recommender systems researchers usually test their hypotheses, novel algorithms, and metrics is by conducting experiments on one or more publicly available datasets of user-item interactions. 
Surveys on the topic of recommender systems research show that the pool of datasets used is small \cite{bobadilla2013recommender}; user behavior data from real-world applications such as media platforms is often proprietary and therefore cannot be used for benchmarking \cite{khusro2016recommender}.
In popularity bias studies, the use of different versions of MovieLens is exceedingly common \cite{wang2023survey}, which leads us to wonder whether studies can be conclusive when they are solely carried out on a few datasets.
In our approach, we include a data synthesis step that allows us to experiment with different data distributions and observe the result. 
Data synthesis is a much-discussed topic in recommender systems research, though it is usually discussed in the context of privacy \cite{slokom2018comparing,tso2006empirical}.

The existence of datasets for training and testing is valuable for the recommender systems community; research on publicly available data is necessary in order to ensure reproducibility \cite{said2014comparative}. 
However, as noted by \cite{cremonesi2021progress}, sharing the used data is not always sufficient to ensure basic reproducibility. 
Studies showed that in most cases recommender systems papers presented at big conferences did not provide code for their data preprocessing or hyperparameter tuning \cite{ferrari2021troubling,ferrari2019we}.
This is also the case in popularity bias research; studies are often not accompanied by code, and sometimes do not describe the data filtering or hyperparameter setting \cite{abdollahpouri2019unfairness,abdollahpouri2020connection}. 
Therefore, concluding that an algorithm or a dataset is prone to popularity bias becomes challenging, as it is not possible to verify or falsify the claims \cite{cremonesi2021progress}.
To showcase this issue, we experiment with algorithms implemented in different libraries and with different parameter configurations.

\section{Identifying Data Characteristics and Algorithm Configurations}
In this section, we identify data characteristics and algorithm configurations that can influence popularity bias propagation in the context of a rating prediction and top-10 recommendation task.
First, we locate data characteristics that can have an effect on whether popularity bias is propagated, inspired by the functionality of UserKNN.
UserKNN is a relatively simple algorithmic approach that simulates a `word-of-mouth' setting and has lower dependence on non-intuitive parameters that impact optimization (e.g., learning rate).
Accordingly, we form a set of data scenarios that combine the located characteristics.
Second, we inspect UserKNN and locate configurations that can be potentially impactful for popularity bias propagation.
 
\subsection{Data Characteristics}\label{subsec:data_charact}
Whether or not popularity bias is propagated depends on how popularity manifests in the dataset at hand. 
We discuss the relation between rating and popularity and the preferences of influential users.

\paragraph{Relation Between Rating and Popularity}
In the context of a rating prediction task, an algorithm aims to predict a future rating for every user of every item they have not already consumed.
Given that this is done by considering the other users' ratings, it may be that items with high average rating will be prioritized by the system. 
Popularity bias studies often do not disclose whether the popular items in the dataset\footnote{Note that item popularity in this context is equivalent to how many users have interacted with the item in a given dataset.} also have high ratings, but instead assume that their frequent recommendation is solely due to their popularity.

\paragraph{Influential Users}
In the context of UserKNN, certain users may be influential because they neighbour with many other users.
For example, if only two users have rated an item and they have not rated any other items, then this item will not be recommended to anyone, because the two users are not influential at all within the system.
Consequently, it is interesting to investigate the notion of user influence and whether the result is dominated by the preferences of users who, because of their large profile size, are more likely to have many neighbours.

\paragraph{Data Scenarios}
We synthesize data that follows a long-tail distribution for items and users, as it is discussed as a prerequisite for popularity bias to occur \cite{brynjolfsson2006niches,celma2008hits}.
Specifically, we choose the interactions in a subset of the Book-Crossing dataset \cite{naghiaei2022unfairness,ziegler2005improving} as a baseline, but remove the rating values. 
To reflect on the observations above, we form a set of scenarios around the relationship between popularity, rating and user influence to assign a synthesized rating to each interaction.
This approach allows us to simulate a real-world scenario where consumption is long-tail, while still experimenting with data properties relevant for popularity bias.
We recognize that the scenarios are not necessarily realistic.
User tendencies are likely to be more subtle in real world situations.
However, we believe that experimenting with extreme behaviors can help us showcase the effect that we are investigating, and lead the way for more nuanced experimentation.

The scenarios, as well as the process we followed to generate each of them are as follows:
\begin{enumerate}
    \item \textbf{Scenario 1: There is no relation between popularity and rating}: For each interaction, draw a rating value between 1 and 10 uniformly at random.\label{scenario:1}
    \item \textbf{Scenario 2: Popular items are generally rated higher by the users}: For each interaction, draw a rating value between 1 and 10 from a normal distribution, where the mean is the popularity of the item normalized between 1 and 10.\label{scenario:2}
    \item \textbf{Scenario 3: Popular items are generally rated lower by the users}: For each interaction, draw a rating value between 1 and 10 from a normal distribution, where the mean is the opposite of the popularity of the item normalized between 1 and 10.  \label{scenario:3}
    \item \textbf{Scenario 4: Only users with big profiles rate popular items higher}: For each interaction, draw a rating value between 1 and 10 uniformly at random. For the users with the 20\% largest profiles, replace by drawing from a Poisson distribution where the mean is the popularity of the item normalized between 1 and 10.  \label{scenario:4}
    \item\textbf{Scenario 5: Only users with big profiles rate popular items lower}: For each interaction, draw a rating value between 1 and 10 uniformly at random. For the users with the 20\% largest profiles, replace by drawing from a Poisson distribution where the mean is the opposite of the popularity of the item normalized between 1 and 10.\label{scenario:5}
\end{enumerate}

\begin{figure*}[!htbp]
  \centering
 
    \subfloat[Scenario 1]{\includegraphics[width=0.3\textwidth]{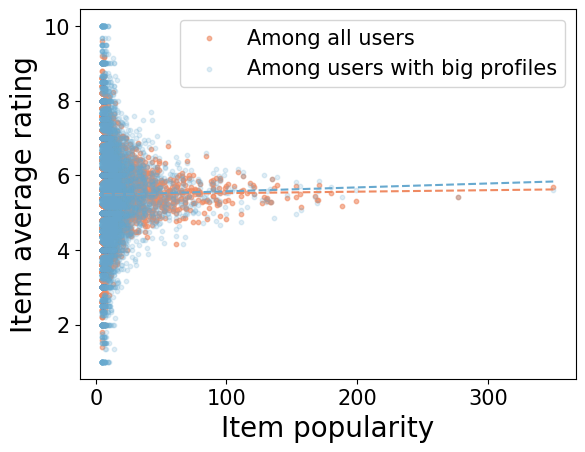}\label{fig:sc1}}
  \subfloat[Scenario 2]{\includegraphics[width=0.3\textwidth]{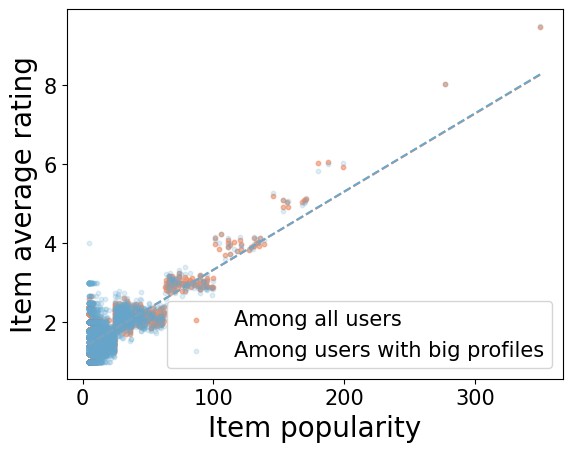}\label{fig:sc2}}
  \subfloat[Scenario 3]{\includegraphics[width=0.3\textwidth]{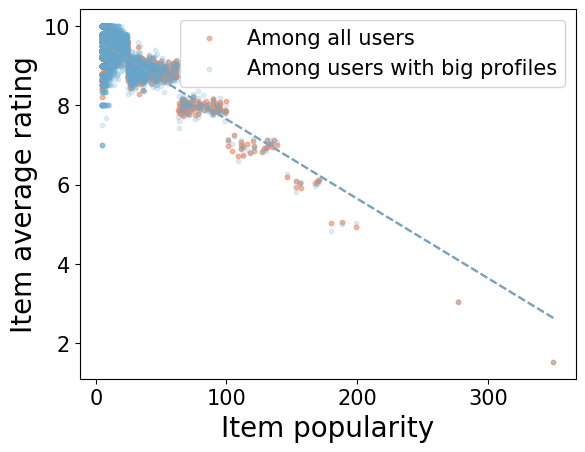}\label{fig:sc3}}

\hspace{0.05\textwidth}
  \subfloat[Scenario 4]{\includegraphics[width=0.3\textwidth]{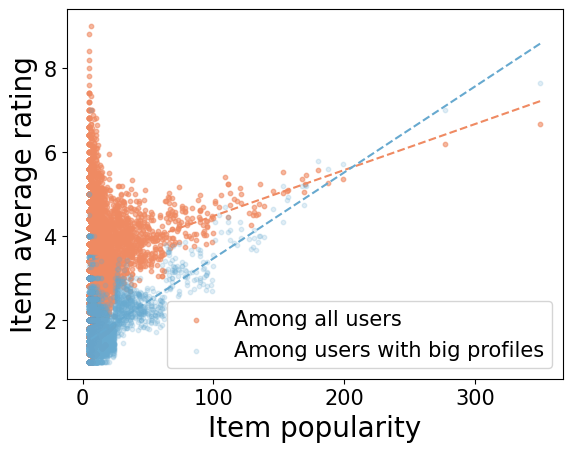}\label{fig:sc4}}
\hspace{0.05\textwidth}
  \subfloat[Scenario 5]{\includegraphics[width=0.3\textwidth]{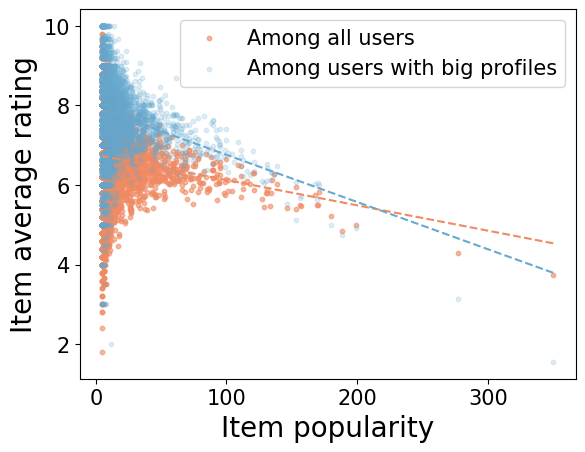}\label{fig:sc5}}
  \hspace{0.05\textwidth}
\hfill
  
  \caption{Relation between item average rating and item popularity among all users and among users with the 20\% largest profiles given five synthetic datasets.}
  \label{fig:data_analysis}
\end{figure*}

Figures \ref{fig:sc1} to \ref{fig:sc5} show the correlation between item average rating and item popularity within the five synthetic datasets.
There is a noticeable effect of the data characteristics discussed in this study, given that the synthetic datasets were constructed with them in mind.
Specifically, scenario 1 shows no relation between average rating and popularity, as the ratings were drawn uniformly at random.
Scenarios 2 and 3 showcase a very positive and a very negative correlation, respectively.
For scenario 4, we see a positive correlation, which is higher for users with large profiles, and for scenario 5 a negative correlation, even lower for users with large profiles.

\subsection{Algorithm Configurations}
In this section, we describe the UserKNN configurations examined.
\subsubsection{UserKNN}\label{subsubsec:userknn}

Despite UserKNN's simplicity, there are configuration choices that can potentially greatly influence the result. 
We identified the following: minimum similarity, the items considered for similarity, and minimum neighbours.

\paragraph{Minimum Similarity} It is common in UserKNN implementations that not all users who rated an item are considered. 
Instead, the notion of neighbourhood is introduced; only the users most similar to the target user are taken into account when producing a predicted score.
The filtering can be done by introducing a cut off value of minimum similarity for consideration, among other techniques.

\paragraph{Items for Similarity} Given a similarity metric (e.g., cosine similarity), a design choice still has to be made on whether similarity between two users will be calculated for their full rating vectors, or only for ratings on items these two users have in common. 
See section 2.3.1.1 in \citet{aggarwal2016recommender} for clarification.

\paragraph{Minimum Neighbours} When the neighbourhood is constructed for a given user, then a score is predicted for each item in a list of candidates. 
In some implementations, the predicted score is not calculated for all potential items.
Instead, the algorithm focuses on items that have been rated by at least a minimum number of neighbours of the current user.

It is worth noting that LensKit and Cornac differ when it comes to these choices as seen in Table \ref{tab:choices}. Two of the parameters tested are not configurable in Cornac, while the third is not configurable in either frameworks.

\begin{table}[htp!]
\centering
\caption{Configuration choices related to UserKNN made by LensKit for Python and Cornac.}

\begin{tabular}{lcc} 
\hline
\textbf{Configuration choice} & \textbf{LensKit for Python} & \textbf{Cornac} \\ \hline

Minimum similarity & Configurable (default: 0) & -1 \\
Items for similarity & All items & Common items \\ 
Minimum neighbours & Configurable (default: 1) & 1 \\ \hline
\end{tabular}
\label{tab:choices}
\end{table}

\section{Experimental Setup}\label{sec:expset}
In this section, we describe the experiments that we run in order to determine the effect of data characteristics and algorithm configuration on the propagation of popularity bias.
The Jupyter notebooks that contain the experiments\footnote{\url{https://github.com/SavvinaDaniil/DiagnosingBiasRecSys/}} have been made open source.
For every synthetic data scenario, we perform a recommendation process given every version of UserKNN.
Specifically, we test the following versions given the configurations discussed in section \ref{subsubsec:userknn}:
\begin{itemize}
    \item Min. similarity 0, over all items, 1 min. neighbour.
    \item Min. similarity 0, over all items, 2 min. neighbours.
    \item Min. similarity -1, over all items, 1 min. neighbour.
    \item Min. similarity -1, over all items, 2 min. neighbours.
    \item Min. similarity -1, over common items, 1 min. neighbour.
\end{itemize}

For each UserKNN version and each dataset, we perform optimization based on RMSE to find the best values for some of the non-fixed hyperparameters of the respective version.
The resulting hyperparameters can be seen in our repository.
Afterwards, we divide the users into training and test users in a 5-fold cross validated way. 
We make sure to use the same splits for all algorithms and all versions.
For every test user, we use 80\% of their ratings for training and the remaining 20\% for testing, which is an option in LensKit. 
We train the model on the training set. 
For each user in the test set, we predict a rating for every item they have not rated in the training set, rank the items based on the predicted score and recommend the top-10 items, in line with recent studies on popularity bias. 

We report on RMSE and NDCG@10 to estimate the effectiveness of the rating prediction and ranking, respectively.
We also calculate the following widely used metrics on the recommended lists to estimate popularity bias propagation:

\begin{enumerate}
    \item \textbf{Popularity Correlation (PopCorr)}: The correlation between popularity in training set and recommendation frequency for every item.
    \item \textbf{Average Recommendation Popularity (ARP)}: The average popularity of the items in the recommended lists \cite{yin2012challenging,abdollahpouri2019managing}. 
    \item \textbf{Popularity Lift (PL)}: The average relative difference in popularity between the recommended items and the items in the users' profiles \cite{abdollahpouri2020connection}.

\end{enumerate}

Finally, for every dataset we perform a Mann–Whitney $U$ test to observe whether there is a significant difference among configurations for ARP and PL, and include the result in the respective tables.

\section{Results}
 In this section, we provide insights into how UserKNN configurations impact popularity bias and performance for the different datasets by presenting the results across the set of metrics listed in section \ref{sec:expset} in table \ref{tab:userknn}.

\begin{table}[htp!]
    \centering
    \caption{Popularity bias and performance of different UserKNN configurations given different data scenarios. OverCommon set to True corresponds to the Cornac implementation, and set to False to the LensKit implementation. For the popularity bias metrics, we embolden the highest value among configurations. For ARP and PL, we use the asterisk (*) to signify which values are significantly lower than the highest one according to a Mann–Whitney $U$ test with $p<0.005$.}
    \begin{tabular}{lllllrrHrr}

\toprule
             & &       &      &  Pop &    ARP$\uparrow$ &       PL$\uparrow$ &  AggDiv$\downarrow$ &    RMSE$\downarrow$ &NDCG  \\
Data & Min & Over & Min &     Corr$\uparrow$      &        &          &         &        & @10$\uparrow$         \\
 Scenario& Sim & Common & Nbrs &           &        &          &         &        &          \\
 
\midrule
\midrule
Scenario 1 & -1 & False & 1 &    0.018 &  0.002* &  -32.285* &   0.400 &  3.502 &    0.001 \\
           &   &       & 2 &    0.418 &  0.004* &   21.252* &   0.681 &  3.352 &    0.003 \\
           &   & True & 1 &    0.004 &  0.002* &  -35.746* &   0.400 &  3.337 &    0.001 \\
           & 0 & False & 1 &    0.101 &  0.003* &  -12.827* &   0.340 &  3.624 &    0.002 \\
           &   &       & 2 &    \textbf{0.615} &  \textbf{0.005} &   \textbf{65.440} &   0.595 &  3.464 &    0.005 \\
\hline
Scenario 2 & -1 & False & 1 &    0.596 &  0.021* &  426.621* &   0.197 &  1.188 &    0.019 \\
           &   &       & 2 &    \textbf{0.614} &  0.022* &  447.618* &   0.339 &  1.190 &    0.021 \\
           &   & True & 1 &    0.604 &  0.015* &  305.197* &   0.238 &  1.150 &    0.013 \\
           & 0 & False & 1 &    0.552 &  \textbf{0.027} &  \textbf{632.300} &   0.080 &  1.040 &    0.023 \\
           &   &       & 2 &    0.562 &  0.027 &  591.966 &   0.173 &  1.026 &    0.025 \\
           \hline
Scenario 3 & -1 & False & 1 &    0.559 &  0.008* &  187.197* &   0.263 &  1.182 &    0.002 \\
           &   &       & 2 &    \textbf{0.728} &  \textbf{0.008} &  \textbf{192.127} &   0.485 &  1.182 &    0.002 \\
           &   & True & 1 &    0.522 &  0.006* &  151.686* &   0.290 &  1.151 &    0.001 \\
           & 0 & False & 1 &    0.025 &  0.002* &  -35.765* &   0.512 &  1.044 &    0.001 \\
           &   &       & 2 &    0.161 &  0.003* &  -13.100* &   0.688 &  1.034 &    0.004 \\
           \hline
Scenario 4 & -1 & False & 1 &    0.253 &  0.003* &   23.063* &   0.488 &  2.502 &    0.001 \\
           &   &       & 2 &    \textbf{0.772} &  0.006* &   97.490* &   0.681 &  2.404 &    0.004 \\
           &   & True & 1 &    0.184 &  0.003* &    8.669* &   0.511 &  2.458 &    0.001 \\
           & 0 & False & 1 &    0.588 &  0.008* &  164.549* &   0.350 &  2.500 &    0.004 \\
           &   &       & 2 &    0.701 &  \textbf{0.014} &  \textbf{297.047} &   0.505 &  2.386 &    0.010 \\
           \hline
Scenario 5 & -1 & False & 1 &    0.087 &  0.003* &   -7.924* &   0.422 &  2.880 &    0.001 \\
           &   &       & 2 &    \textbf{0.623} &  0.005* &   \textbf{57.969} &   0.679 &  2.776 &    0.003 \\
           &   & True & 1 &    0.057 &  0.002* &  -16.243* &   0.433 &  2.783 &    0.001 \\
           & 0 & False & 1 &    0.136 &  0.003* &  -16.122* &   0.506 &  2.914 &    0.003 \\
           &   &       & 2 &    0.612 &  \textbf{0.005} &   42.849 &   0.674 &  2.794 &    0.006 \\
\bottomrule
\bottomrule
\end{tabular}
    
    \label{tab:userknn}
\end{table}

Performance varies across the data scenarios.
RMSE specifically is lower for scenarios \ref{scenario:2} and \ref{scenario:3} compared to the other three.
In these two scenarios, users tend to agree between them on whether they like popular items or not, which facilitates the rating prediction task.
NDCG@10 is the highest for scenario \ref{scenario:2}, where popular items are highly rated by the users. 
In this case, the rating prediction and ranking tasks are linked, since the highest ranked (i.e., popular items) are also highly rated.

Popularity bias also varies across the data scenarios, and the effect depends on the algorithm configuration.
In the following paragraphs, we describe and reflect on the most impactful effects of the interaction between data and configuration.

For scenario \ref{scenario:1} where ratings are uniformly at random generated, there is no notable popularity bias propagation observed when minimum neighbours are set to 1, while there is bias when minimum neighbours is set to 2.
Increasing minimum neighbours results in higher popularity bias for all datasets and metrics.

In scenario \ref{scenario:3} where all users agree that popular items are bad, popularity bias is not propagated when minimum similarity is set to 0.
However, when setting minimum similarity to -1, we can observe popularity bias propagation across all metrics.
The reason is that users with completely different opinions are considered and their opinions count negatively. 
Therefore, popular items still get recommended since everyone's ``negative" neighbours dislike them, and we can observe popularity bias propagation across all metrics.

When considering only common items to calculate similarity, users with smaller profiles have a larger influence.
This is relevant in scenario \ref{scenario:4} where users with large profiles like popular items.
Table \ref{tab:userknn} shows that even though scenario \ref{scenario:4} still leads to popularity bias, considering only common items reduces it across all metrics.
Therefore, this implementation choice can have a big impact on whether popularity bias is propagated and to what extent.

Finally, the value for minimum neighbours largely influences popularity bias. 
Across almost all scenarios and metrics, increasing minimum neighbours from 1 to 2 leads to increased popularity bias.
By setting a higher neighbour barrier for considering an item for recommendation, it follows that less popular items will be disadvantaged. 
This result is particularly relevant given that the parameter of minimum neighbours could only be tweaked 
in one of the considered frameworks.
Thus, 
studies that use Cornac or LensKit might reach different conclusions on the extent of popularity bias propagated by UserKNN.

\emph{The above results show that configuration has a significant effect on whether or not popularity bias is observed. All three UserKNN configuration choices affect the observed bias. This effect is different for different data scenarios, with some configurations leading to low popularity bias on one dataset while leading to high popularity bias on another dataset. }

\section{Discussion}
\subsubsection*{Implications of the present study}
Our research shows that multiple data and configuration factors can have an effect on whether bias is propagated.
Relying on frameworks readily available to researchers is convenient and a concrete step towards reproducibility, but requires being aware and detailed about the limitations.
When simple parameters, such as minimum neighbours in UserKNN, are so influential, it raises questions on how generalizable research on recommender systems bias can be.
Our results indicate that bias studies can only draw conclusions within the limits of their specific research and not further than that.
It follows that being explicit about the context within which a type of bias is studied is crucial, both in terms of data characteristics and implementation.
It is a known issue in recommender systems literature that implementation details are often not disclosed by studies.
Even in cases where they are, guessing the effect of different hyperparameters that are not present in an implementation or experimented with is not trivial. 
Bias reporting is definitely not complete if it is not accompanied by clarity around the characteristics, goals and limitations of the system that is being studied.

\subsubsection*{Recommendations of the present study}
Based on the results of the present study, we have come up with two recommendations towards researchers who study bias in recommender systems.

First, researchers should analyze and report on the dataset characteristics that might impact the type of bias they are concerned with. 
For UserKNN, the relationship between rating and popularity, as well as the preferences of users with large profiles impact popularity bias and should be taken into account in relevant studies. 
For other algorithms and types of bias, there could be other relevant characteristics. 
Such analysis will help the reader understand the extent to which the results are a result of the dataset characteristics.

Second, researchers should test multiple algorithm configurations when measuring bias propagation.
In a similar way that the community expects x-fold cross validation, since presenting the results of only one run may not be reliable, we could expect to see results on multiple algorithm configurations as well.
If the conclusions are only valid for one specific configuration of the algorithm at hand, then that should be clear in the limitations of the study.

\subsubsection*{Limitations of the present study}

Despite our extensive testing, the results are potentially sensitive to our own experimental design, such as the method for train-test splitting or randomness in the data generation process.
Similarly, instantiating the two UserKNN implementations with the exact same configuration choices is not possible due to some of the parameters not being configurable.
As a result, there might be implementation differences between the frameworks that we are not aware of and cause part of the variation in results, irrespectively of the configurations tested.
However, these observations highlight the importance of this line of research instead of hindering it, since they hint that data and implementation dependence might be present more often than we think.
Additionally, we recognize that the scholar community has generally moved on from explicit user preferences.
We do not focus on implicit feedback given that recent studies on popularity bias are often performed on datasets with explicit ratings in the context of rating prediction tasks \cite{abdollahpouri2019unfairness,abdollahpouri2020connection,kowald2020unfairness}.
Our goal is to show that conclusions in literature on the topic of popularity bias propagation can be volatile and require further scrutiny when it comes to their dependence on implementation and data domain. 
Future work can focus on components that we chose not to investigate in this study, such as more advanced algorithms and other open libraries for implementation.
Further nuance can be introduced in the data synthesis part, by allowing for more complex relationships between popularity, rating and user influence.
Finally, a similar approach could be used to investigate other known biases in recommender systems.

\section{Conclusion}
In this study, we reflected on the need for fundamental understanding of the relationship between data, algorithms and bias in recommender systems. 
We focused on reporting on popularity bias propagated by UserKNN , and tracked configurations and data characteristics that are of importance in its propagation. 
Accordingly, we generated a set of synthetic datasets, experimented with performing a recommendation process on them using different configurations of UserKNN, and evaluated popularity bias using well-known metrics.
We found that even when the distribution of popularity in the dataset is long-tail, popularity bias is not unavoidable.
We showed that the relationship between popularity and rating, as well as the preferences of users with big profiles largely impacts bias.
We highlighted the sensitivity of bias propagation to algorithm configuration and, by extension, framework implementation.
Our observations point to methodology and reproducibility issues that extend further than a specific use case, to the recommender systems field at large.

Recommender systems are widely used in our online lives, and bias propagation by such systems can have serious societal impact.
With this work, we hope to have called attention to the ambiguity in bias reporting and motivated researchers to strive for reproducibility and highlight specificity when appropriate.
\bibliographystyle{ACM-Reference-Format}
\bibliography{main}

\end{document}